\documentclass[11pt]{article}
\usepackage{axodraw}
\usepackage{epsfig}
\usepackage{amsfonts}
\usepackage{amsmath}
\usepackage{bbm}
\usepackage{cite}
\input pix.sty

\allowdisplaybreaks[1]
\def\undertilde#1{\mathop{\vtop{\ialign{##\cr$\textstyle{#1}$\cr%
\noalign{\kern1pt\nointerlineskip}\hfil$\mathchar"0365$\hfil\cr}}}}
\def\wideundertilde#1{\mathop{\vtop{\ialign{##\cr$\textstyle{#1}$\cr%
\noalign{\kern1pt\nointerlineskip}\hfil$\mathchar"0367$\hfil\cr}}}}
\renewcommand{\eq}{eq.~}
\renewcommand{\eqs}{eqs.~}
\renewcommand{\se}{sec.~}

\newcommand{\vc}[1]{{\bf #1}}

\newcommand{\alphas}{\alpha_{\rm s}}
\newcommand{\Nf}{N_{\rm f}}
\newcommand{\Nc}{N_{\rm c}}

\newcommand{\gB}{g_\rmii{B}}

\newcommand{\gammaE}{\gamma_\rmii{E}}

\newcommand{\rmO}{{\mathcal{O}}}
\newcommand{\bmu}{\bar\mu}

\newcommand{\CF}{C_\rmii{F}}

\def\lsi{\raise0.3ex\hbox{$<$\kern-0.75em\raise-1.1ex\hbox{$\sim$}}}
\def\gsi{\raise0.3ex\hbox{$>$\kern-0.75em\raise-1.1ex\hbox{$\sim$}}}

\newcommand{\gsim}{\mathop{\gsi}}

\newcommand{\nB}{n_\rmii{B}}

 \renewcommand{\nB}[1]{n_\rmii{B{#1}}}
\newcommand{\rmii}[1]{{\mbox{\tiny\rm{#1}}}}
\newcommand{\re}{\mathop{\mbox{Re}}}

\newcommand{\Tint}[1]{{\hbox{$\sum$}\!\!\!\!\!\!\!\int\,}_{\!\!\!\!\raise-0.9ex\hbox{$\scriptstyle{#1}$}}}
\newcommand{\Tinti}[1]{{{\Sigma}\!\!\!\!\raise0.3ex\hbox{$\int$}_\rmii{${#1}$}}}

\newcommand{\bi}{\begin{itemize}}
\newcommand{\ei}{\end{itemize}}
\newcommand{\hide}[1]{ }

\def\TAsc(#1,#2)(#3,#4,#5)%
{\SetWidth{2.0}\CArc(#1,#2)(#3,#4,#5)\SetWidth{1.0}}
\def\Lwidth{3}

\def\TAgl(#1,#2)(#3,#4,#5){\SetWidth{2.0}\PhotonArc(#1,#2)(#3,#4,#5){\Lwidth}%
{6.283 #3 mul 360 div #4 #5 sub #4 #5 sub mul sqrt mul Tdensity mul}%
\SetWidth{1.0}}
\def\TLgl(#1,#2)(#3,#4){\SetWidth{2.0}\Photon(#1,#2)(#3,#4){\Lwidth}
{#1 #3 sub #1 #3 sub mul #2 #4 sub #2 #4 sub mul add sqrt Tdensity mul}%
\SetWidth{1.0}}
\def\Lwidth{1.3}

\newcommand{\picu}[1]{\;\parbox[c]{30pt}{\begin{picture}(30,30)(0,0)
\SetWidth{1.0}\SetScale{1.0} #1 \end{picture}}\; }
\def\EleA{\picu{%
 \CArc(15,15)(15,0,360)%
 \Lgl(15,0)(15,30)%
 \COval(15,0)(2,2)(0){Black}{Black}%
 \COval(15,30)(2,2)(0){Black}{Black}%
}}
\def\EleAp{\picu{%
 \CArc(15,15)(15,0,360)%
 \Lgl(15,0)(15,30)%
 \COval(15,0)(2,2)(0){Black}{Black}%
 \COval(15,30)(2,2)(0){Black}{Black}%
 \Agl(15,35)(5,0,360)%
}}
\def\EleB{\picu{%
 \CArc(15,15)(15,0,360)%
 \Lgl(15,0)(15,30)%
 \COval(15,0)(2,2)(0){Black}{Black}%
 \COval(15,30)(2,2)(0){Black}{Black}%
 \Agl(29,15)(8,100,260)%
}}
\def\EleBnorm{\piccb{%
 \CArc(15,15)(15,0,360)%
 \Lgl(15,0)(15,30)%
 \COval(15,0)(2,2)(0){Black}{Black}%
 \COval(15,30)(2,2)(0){Black}{Black}%
 \CArc(60,15)(15,0,360)%
 \Agl(74,15)(8,100,260)%
 \Line(30,0)(45,30)%
 \Line(-3,0)(-3,30)%
 \Line(-3,30)(2,30)%
 \Line(-3,0)(2,0)%
 \Line(78,0)(78,30)%
 \Line(78,30)(73,30)%
 \Line(78,0)(73,0)%
}}
\def\EleBB{\picu{%
 \CArc(15,15)(15,0,360)%
 \Lgl(15,0)(15,30)%
 \Lgl(0,15)(12,15)%
 \Lgl(18,15)(30,15)%
 \COval(15,0)(2,2)(0){Black}{Black}%
 \COval(15,30)(2,2)(0){Black}{Black}%
}}
\def\EleC{\picu{%
 \CArc(15,15)(15,0,360)%
 \Agl(7,4)(8,-30,130)%
 \Agl(23,26)(8,150,310)%
 \COval(15,0)(2,2)(0){Black}{Black}%
 \COval(15,30)(2,2)(0){Black}{Black}%
}}
\def\EleD{\picu{%
 \CArc(15,15)(15,0,360)%
 \Agl(23,4)(8,50,210)%
 \Agl(23,26)(8,150,310)%
 \COval(15,0)(2,2)(0){Black}{Black}%
 \COval(15,30)(2,2)(0){Black}{Black}%
}}
\def\EleE{\picu{%
 \CArc(15,15)(15,0,360)%
 \Agl(40,30)(25,180,242)%
 \Agl(40,0)(25,118,137)%
 \Agl(40,0)(25,150,180)%
 \COval(15,0)(2,2)(0){Black}{Black}%
 \COval(15,30)(2,2)(0){Black}{Black}%
}}
\def\EleF{\picu{%
 \CArc(15,15)(15,0,360)%
 \Lgl(15,0)(15,30)%
 \Agl(40,40)(26,205,245)%
 \COval(15,0)(2,2)(0){Black}{Black}%
 \COval(15,30)(2,2)(0){Black}{Black}%
}}
\def\EleG{\picu{%
 \CArc(15,15)(15,0,360)%
 \Agl(40,15)(30,150,210)%
 \Agl(-10,15)(30,-30,30)%
 \COval(15,0)(2,2)(0){Black}{Black}%
 \COval(15,30)(2,2)(0){Black}{Black}%
}}
\def\EleGp{\picu{%
 \CArc(15,15)(15,0,360)%
 \Agl(15,25)(5,0,360)%
 \Agl(15,5)(5,0,360)%
 \COval(15,0)(2,2)(0){Black}{Black}%
 \COval(15,30)(2,2)(0){Black}{Black}%
}}
\def\EleH{\picu{%
 \CArc(15,15)(15,0,360)%
 \Lgl(15,0)(15,12)%
 \Agl(10,21.5)(10,-65,65)%
 \Agl(20,21.5)(10,115,245)%
 \COval(15,0)(2,2)(0){Black}{Black}%
 \COval(15,30)(2,2)(0){Black}{Black}%
}}
\def\EleI{\picu{%
 \CArc(15,15)(15,0,360)%
 \Lgl(15,0)(15,30)%
 \Lgl(15,15)(30,15)%
 \COval(15,0)(2,2)(0){Black}{Black}%
 \COval(15,30)(2,2)(0){Black}{Black}%
}}
\def\EleJ{\picu{%
 \CArc(15,15)(15,0,360)%
 \Lgl(15,0)(15,30)%
 \COval(15,0)(2,2)(0){Black}{Black}%
 \COval(15,30)(2,2)(0){Black}{Black}%
 \GCirc(15,15){4}{0.5}
}}
\newcommand{\piG}[1]{\;\parbox[c]{185pt}{\begin{picture}(10,40)(-175,-15)
\SetWidth{1.0}\SetScale{1.0} #1 \end{picture}}\;}
\def\LattGE{\piG{%
 \SetWidth{1.0} 
 \Line(-130,10)(-110,10)%
 \Line(-100,10)(-100,20)\Line(-100,20)(-90,20)%
 \Line(-85,10)(-75,10)\Line(-75,10)(-75,20)%
 \Line(-65,20)(-45,20)%
 \Line(-35,20)(-35,10)\Line(-35,10)(-25,10)%
 \Line(-20,20)(-10,20)\Line(-10,20)(-10,10)%
 \Vertex(-130,10){1}
 \Vertex(-110,10){1}
 \Vertex(-100,10){1}
 \Vertex(-90,20){1}
 \Vertex(-85,10){1}
 \Vertex(-75,20){1}
 \Vertex(-65,20){1}
 \Vertex(-45,20){1}
 \Vertex(-35,20){1}
 \Vertex(-25,10){1}
 \Vertex(-20,20){1}
 \Vertex(-10,10){1}
 \Text(-87.5,15)[c]{$-$}
 \Text(-22.5,15)[c]{$-$}
 \Text(-135,15)[c]{$\bigl\langle\,$}
 \Text(-105,15)[c]{$\bigl($}
 \Text(-70,15)[c]{$\bigr)$}
 \Text(-40,15)[c]{$\bigl($}
 \Text(-5,15)[c]{$\bigr)$}
 \Text(0,15)[c]{$\bigr\rangle$}
 \Text(-138,15)[r]{$\sum_{i=1}^{3}\re\tr$}
 \Line(-190,3)(5,3)
 \Text(-110,-10)[r]{$- 3 a^4 \re\tr\bigl\langle$}
 \Line(-105,-10)(-35,-10)
 \Vertex(-105,-10){1}
 \Vertex(-35,-10){1}
 \Text(-30,-10)[c]{$\bigr\rangle$}
 \Text(-195,2)[r]{$G^{ }_{\rmii{E},\rmi{latt}}(\tau) \; \equiv \; $}
  }}
\def\numerator{\pis{%
 \Text(10,10)[c]{$\displaystyle
  T \sum_{q_n'} \frac{ \hphantom{\undertilde{q_n} e^{i q_n \tau}} 
  }{\tilde{q}_n}
  $}%
 \Text(23,21)[c]{$ \undertilde{q_n} e^{i q_n \tau}
  $}
 }}
\makeatletter \@addtoreset{equation}{section} \makeatother
\renewcommand{\theequation}{\arabic{section}.\arabic{equation}}
\makeatletter
\renewcommand\section{\@startsection {section}{1}{\z@}%
                                   {-5.5ex \@plus -1ex \@minus -.2ex}% bfr-
                                   {2.3ex \@plus.2ex}%
                                   {\normalfont\large\bfseries}}
\renewcommand\subsection{\@startsection{subsection}{2}{\z@}%
                                     {-3.25ex\@plus -1ex \@minus -.2ex}%
                                     {1.5ex \@plus .2ex}%
                                     {\normalfont\normalsize\bfseries}}
\renewcommand\thesection {\@arabic\c@section}
\renewcommand\thesubsection   {\thesection.\@arabic\c@subsection}
\renewcommand{\@seccntformat}[1]{%
\csname the#1\endcsname.\hspace{1.0em}}
\makeatother
\begin{document}

\flushbottom

\begin{titlepage}

\begin{flushright}
% OUTLINE  \\ 
% DRAFT M.L. \\ 
% arXiv:1601.01573\\ 
February 2016 \\ 
\vspace*{1cm}
\end{flushright}

\begin{centering}
\vfill

{\Large{\bf
 Perturbative renormalization of the electric field correlator
}} 

\vspace{0.8cm}

C.~Christensen and %$^{\rm a}$, %%
M.~Laine % $^{\rm a}$, %%\footnote{laine@itp.unibe.ch}

\vspace{0.8cm}
 
% $^\rmi{a}$%
{\em
AEC, Institute for Theoretical Physics, University of Bern, \\  
Sidlerstrasse 5, CH-3012 Bern, Switzerland
\\}

\vspace*{0.8cm}

\mbox{\bf Abstract}
 
\end{centering}
 
\vspace*{0.3cm}
 
\noindent
The momentum diffusion coefficient of a heavy quark in a hot QCD 
plasma can be extracted as a transport coefficient related to the 
correlator of two colour-electric fields dressing a Polyakov loop. 
We determine the perturbative renormalization factor for a particular 
lattice discretization of this correlator within Wilson's SU(3) gauge 
theory, finding a $\sim 12$\% NLO correction for values of the bare 
coupling used in the current generation of simulations. The impact of this 
result on existing lattice determinations is commented upon, and 
a possibility for non-perturbative renormalization through 
the gradient flow is pointed out. 

\vfill

%% %\noindent
%% %PACS numbers: 
%% %11.10.Wx, %        Finite temperature field theory
%% { %11.15.Ha, %        Lattice gauge theory } 
%% %12.38.Bx, %        Perturbative calculations in QCD
%% %12.38.Mh, %        Quark--gluon plasma
%% %14.40.Nd, %        Bottom mesons
%% %\\
%% %Keywords:
%%  Thermal field theory; lattice gauge theory; quark-gluon plasma; 
%%    heavy quark physics; renormalization; perturbative QCD
 
\vspace*{1cm}
  
% \noindent
% date

\vfill

\end{titlepage}

%%%%%%%%%%%%%%%%%%%%%%%%%%% SECTION %%%%%%%%%%%%%%%%%%%%%%%%%%%%%%%%%%%%%
%
\section{Introduction}

If a system in thermodynamic equilibrium is displaced slightly by means
of some external perturbation, it tends to relax back to equilibrium. The 
rate at which this happens depends on the nature of the perturbation. 
Among the simplest perturbations is that the current $J^{\mu}$
associated with some conserved particle species is not aligned with 
the flow velocity $u^{\mu}$ of the heat bath: 
$\langle (\eta^{\mu\nu} - u^{\mu}u^{\nu})J_\nu \rangle \neq 0$. 
Then the relaxation rate is called the kinetic equilibration rate; 
more generally it can be a function of the wave vector associated
with the perturbation of $J^\mu$, with the rate-proper defined
through the long-wavelength limit. 

In a hot QCD plasma produced in 
heavy ion collision experiments, with a lifetime of $\sim 10$~fm/c 
and a temperature of $\sim 300$~MeV, heavy (charm and bottom) 
quarks provide examples of conserved particle
species. Indeed their weak decays are too slow by many orders of 
magnitude to play a role. 
Moreover the particle and antiparticle number densities are to a good 
approximation 
conserved separately, given that pair creations and annihilations are 
also too slow to take place in a typical event, 
unless $T \gsim 600$~MeV~\cite{chem}. 

Heavy quarks are naturally displaced from kinetic equilibrium, given that
they are generated in an initial hard process which has no knowledge of 
the thermal state and the flow that it develops during 
$\sim 1$~fm/c. It is empirically observed, however, 
that the process of kinetic 
equilibration takes place during the fireball expansion: 
heavy quark jets get quenched, and eventually heavy quarks
participate in hydrodynamic flow almost as efficiently as light quarks do
(cf.,\ e.g.,\ refs.~\cite{alice,phenix,star}). This calls for a theoretical 
computation of their kinetic equilibration rate~\cite{mt}, which can 
subsequently be inserted into phenomenological models permitting 
for comparisons with data 
(cf.\ ref.~\cite{ab} for a review).  

A large body of theoretical and model-based work has already been carried out 
on heavy quark kinetic equilibration. Focussing here on the deconfined phase, 
it has been observed that next-to-leading order (NLO) corrections are 
large and positive~\cite{chm1,chm2}, and that the AdS/CFT setup suggests 
rapid equilibration~\cite{ads1,ads2,ct}. However ultimately 
the problem should be studied with methods of lattice QCD. It turns out that
in the heavy-quark limit, applicable at least for the bottom quarks, 
the lattice challenge appears to be relatively manageable, because
the observable in question reduces to 
a purely gluonic ``electric field correlator''
(cf.\ \eq\nr{GE_def}) whose associated 
spectral function is expected to be smoother than for almost any
other transport observable~\cite{eucl,mink}. Indeed first lattice
measurements making use of this observation have been carried out 
within quenched QCD~\cite{latt_a,kappaE,latt_c,kappa}.

There are several important issues which the existing simulations 
have not solved conclusively. One is taking the continuum limit: 
for a systematic extrapolation the electric field correlator 
requires a finite renormalization factor, 
which should ultimately be determined non-perturbatively. 
Another is analytic continuation: 
after the continuum limit has been taken, a short-distance singularity
can in principle be subtracted~\cite{cond} and the result 
subsequently subjected to a well-defined
analytic continuation algorithm~\cite{cuniberti}, however
implementing this in practice requires exquisite 
statistical precision~\cite{analytic}, so that in the existing studies
theoretically motivated ans\"atze 
have been employed for extracting 
the transport coefficient~\cite{kappaE,latt_c,kappa}. 

The issue addressed in the present paper is that of renormalization. 
Specifically we compute the renormalization factor for 
the electric field correlator at 1-loop order in lattice perturbation
theory~\cite{lpt1,lpt2}, and briefly comment on the possibilities
for non-perturbative renormalization. With such ingredients {\em and} 
improved statistical precision, the current order-of-magnitude
estimate~\cite{kappa} can hopefully be promoted towards 
a quantitative level. 

The plan of this paper is the following. 
After defining the basic observable in \se\ref{se:defs}, 
the technical implementation of the computation is outlined
in \se\ref{se:outline}. Results for individual Feynman diagrams
are documented in \se\ref{se:diags}. The results are put together
in \se\ref{se:ZE} for our final expression for the perturbative
renormalization factor. We offer an outlook and comments
on non-perturbative renormalization in \se\ref{se:concl}.

%%%%%%%%%%%%%%%%%%%%%%%%%%% SECTION %%%%%%%%%%%%%%%%%%%%%%%%%%%%%%%%%%%%%
%
\section{Basic definitions}
\la{se:defs}

The physical observable  underlying our considerations 
is the 2-point correlator
of the time derivatives of the spatial components of the conserved vector 
current, evaluated in an ensemble at a finite temperature $T$. 
After taking the heavy quark-mass limit~\cite{eucl}, the 
correlator reduces to the 2-point function 
of coloured Lorentz forces~\cite{ct}.

We first define the correlator in continuum notation. Letting 
$ 
  U(\tau_b,\tau_a) 
$
be a temporal Wilson line at the spatial position
$\vec{r}\equiv\vec{0}$ and defining colour-electric fields through 
\be
 \gB E_i \equiv i [D_0, D_i] 
 \;, 
\ee
where $D_\mu = \partial_\mu + i \gB A_\mu$ is a covariant derivative
and $\gB$ is the bare gauge coupling of a dimensionally regularized
SU($\Nc$) gauge theory, 
the electric field correlator reads~\cite{eucl}
\be
 G^{ }_{\rmii{E},\rmi{cont}}(\tau) 
 \; \equiv \;
 - \fr13 \sum_{i=1}^{3-2\epsilon} 
 \frac{
  \Bigl\langle
   \re\tr \Bigl[
      U(\beta,\tau) \, 
      \gB E_i(\tau,\vec{0}) \, U(\tau,0) \, 
      \gB E_i(0,\vec{0})
   \Bigr] 
  \Bigr\rangle
 }{
 \Bigl\langle
   \re\tr [U(\beta,0)] 
 \Bigr\rangle
 }
 \;. \la{GE_def}
\ee
Here $\beta\equiv 1/T$ denotes the inverse temperature. 
The global Z($\Nc$) symmetry of the SU($\Nc$) gauge theory 
is assumed to be spontaneously or explicitly broken, so 
that the denominator of \eq\nr{GE_def} is non-zero. 

The lattice 
discretization of $G^{ }_\rmii{E}$ is not unique; we have in 
mind the proposal from ref.~\cite{eucl} which was argued to
have small discretization effects, {\em viz.}
\ba
 && \hspace*{0.5cm} \LattGE \;, 
 \la{GE_lattice}
\ea
where lines indicate parallel transporters, and $a$ is the lattice spacing. 
We focus on pure SU($\Nc$) gauge theory in the following, 
setting $\Nf = 0$. In this case the Z($\Nc$) symmetry is spontaneously
broken in the high-temperature phase. 

Even though the correlator is finite
after coupling constant renormalization~\cite{eucl,rhoE}, 
the continuum and lattice-regularized expressions do differ
by a finite factor in every order of perturbation theory. In the following
we determine this factor at NLO, i.e.\
at $\rmO(\alphas)$. Numerical measurement should be multiplied
by this factor (cf.\ \eq\nr{ZE_def})
in order to obtain the continuum result.
If the spectral function ($\rho^{ }_{\rmii{E},\rmi{cont}}$) 
corresponding to the 
continuum result can subsequently
be extracted, then the corresponding
transport coefficient is known as the ``momentum diffusion coefficient'': 
\be
 \kappa = \lim_{\omega \to 0} 
 \frac{2 T \rho^{ }_{\rmii{E},\rmi{cont}}(\omega)}{\omega}
 \;. 
\ee
The kinetic equilibration rate 
(also known as the drag coefficient $\eta^{ }_D$) is given by
\be
 \Gamma_\rmi{kin} = \frac{\kappa}{2 M_\rmi{kin} T}
 \;, 
\ee
where $M_\rmi{kin}$ is a so-called kinetic mass of the heavy quarks. 
The single factor $\kappa$ yields two different kinetic equilibration rates 
of phenomenological interest, one for the charm and another for bottom quarks. 
In the charm case $1/\Gamma_\rmi{kin}$ could conceivably be as small as 
a few fm/c~\cite{kappa}, which could then offer for a partial qualitative 
explanation for the experimentally observed hydrodynamic 
flow observed in $D$ meson spectra after hadronization. 

%%%%%%%%%%%%%%%%%%%%%%%%%%% SECTION %%%%%%%%%%%%%%%%%%%%%%%%%%%%%%%%%%
%
\section{Technical outline}
\la{se:outline}

Denoting by $U_\mu(X)$ a link matrix which changes in gauge transformations
as 
\be
 U_\mu(X) \to G(X)\, U_\mu(X)\, G^{-1}(X + a \hat{\mu})
 \;, 
\ee
where $\hat{\mu}$ is a unit vector in the $\mu$-direction, 
the observable of \eq\nr{GE_lattice} can be expressed as\footnote{%
 For convenience the imaginary time direction has been reversed 
 with respect to \eq\nr{GE_def}. 
 } 
\be
 G^{ }_{\rmii{E},\rmi{latt}}(\tau) = 
 -\frac{1}{3a^4} \frac{\sum_i \re \tr \langle A\,B\,C\,D \rangle}
 {\re\tr \langle P \rangle}
 \;,  \la{GE_latt}
\ee
where 
\ba
 A & = & U^{ }_0(0)U^{ }_i(0 + a \hat{0})
       - U^{ }_i(0) U^{ }_0(0 + a \hat{i}) \;, \\
 B & = & U^{ }_0(0 + a \hat{0}+a\hat{i}) \cdots 
         U^{ }_0(0 + (\tau - a) \hat{0}+a\hat{i}) \;, \\ 
 C & = & U^{ }_0(0 + \tau\hat{0} + a \hat{i})
         U_i^\dagger(0 + (\tau + a) \hat{0}) 
       - U_i^\dagger(0 + \tau\hat{0} )
         U^{ }_0(0 + \tau \hat{0}) \;, \\
 D & = & U^{ }_0(0 + (\tau + a) \hat{0} ) \cdots 
         U^{ }_0(0 + (\beta - a) \hat{0} ) \;, \\
 P & = & U^{ }_0(0) \cdots 
         U^{ }_0(0 + (\beta - a) \hat{0} ) \;.
\ea
For a perturbative analysis, the link matrices are expressed in terms of
gauge potentials as 
\be
 U_\mu(X) = e^{i a g_0 T^a A^a_\mu (X)}
 \;, 
\ee
where $g_0$ is the bare lattice gauge coupling and $T^a$ are Hermitean
generators of SU($\Nc$), normalized as $\tr(T^a T^b) = \delta^{ab}/2$. 
The gauge potentials can be Fourier-represented as
\be
 A^a_\mu(X) = \Tint{K} A^a_\mu(K) e^{i K\cdot (X+ \frac{a \hat{\mu}}{2})}
 \;, 
\ee
where the sum-integration measure is as defined in \eq\nr{measure2}. 
We use the standard Wilson action, 
with well-known momentum-space Feynman rules
(cf.\ refs.~\cite{lpt1,lpt2}). 

As mentioned before, \eq\nr{GE_lattice} is
unambiguous only when the Z($\Nc$) center symmetry
is explicitly or spontaneously broken.
We have carried out the renormalization computation at 
a high temperature, when the latter is the case. 
Even though formally justifying the use
of the weak-coupling expansion, the finite temperature
has the side effect that 
Matsubara zero modes play a prominent role; indeed, they lead to 
effects of $\rmO(1/a)$ or $\rmO(a)$ in many individual terms. We have
verified the cancellation of these structures, which then also serves as 
a partial crosscheck of our computation. 

%%%%%%%%%%%%%%%%%%%%%%%%%%% SECTION %%%%%%%%%%%%%%%%%%%%%%%%%%%%%%%%%%
%
\section{Results for individual diagrams}
\la{se:diags}

As an illustration of the computational procedure, 
we give in this section results for the individual Feynman diagrams, both 
in dimensional regularization and in lattice regularization. 
In each case only those terms which differ in the 
two schemes are shown; the full results for dimensional regularization, 
including the scheme-independent parts, can be found in ref.~\cite{rhoE}. 
For simplicity the calculations were carried out in Feynman gauge; their sum
is gauge-independent. 

In the expressions below, the common sum-integration ``measure'' 
\be
 \int_\Omega \equiv 
 - \frac{g^2\CF}{3}\sum_{i=1}^{D-1} \Tint{K} \frac{e^{i k_n \tau}}{K^2}
 \la{measure}
\ee
is suppressed. Here $i$ enumerates the spatial directions, $D$
is the space-time dimension, and 
$K = (k_n ,\vec{k})$ is an imaginary-time four-momentum, with 
$k_n$ denoting a Matsubara frequency. The momentum variables
have been so chosen that the components of $K$ can be assumed
small compared with the ultraviolet cutoff, i.e.\ 
$|K| \ll \frac{1}{a}$, where $a$ denotes the lattice spacing. 
The renormalized gauge coupling
of the $\msbar$ scheme is denoted by $g^2$, 
and $\CF \equiv (\Nc^2-1)/(2\Nc)$. 

The contributions of the different Feynman diagrams read as follows. 
A renormalization contribution is obtained by expressing the bare 
continuum and lattice couplings
in terms of the renormalized $\msbar$ coupling.\footnote{% 
 We find it convenient to first present the results in terms of the 
 renormalized $\msbar$ coupling $g^2$, 
 even though at the end of the computation
 $g^2$ will be re-expressed through the bare lattice coupling $g_0^2$. 
 }
For the lattice coupling this 
relation was determined in ref.~\cite{hh} and has
been extended up to 2-loop level in ref.~\cite{lw}. For our purposes, 
\eq(1.1) of ref.~\cite{lw} is best re-expressed as 
\be
 g_0^2 = g^2 + \frac{g^4\Nc}{(4\pi)^2} 
 \biggl[ - \frac{4\pi d_1(a\bmu)}{\Nc} \biggr] + \rmO(g^6)
 \;, \la{L_MS}
\ee
where 
\be
  - \frac{4\pi d_1(a\bmu)}{\Nc} = \frac{11}{3} \ln (a^2\bmu^2) 
  + \fr13 - \pi^2 \biggl(1 + \frac{20 P^{ }_1}{9} + \frac{176 P^{ }_2}{3}
  - \frac{2}{\Nc^2}
 \biggr)
 \;, \la{L_MS_expl}
\ee
and $\bmu$ % $\bmu^2 = 4 \pi e^{-\gammaE} \mu^2$ 
is the scale parameter of the $\msbar$ scheme.
The coefficients $P^{ }_1,P^{ }_2$ read
\ba
 P^{ }_1 & \equiv & \int_Q \frac{a^2}{\tilde{Q}^2}
  =  0.15493339023106021(1) \;, \la{defP1} \\ 
 P^{ }_2 & \equiv & 
 \lim_{m\to 0} \biggl[ 
   \frac{\ln(a^2m^2)}{(4\pi)^2}
   + \int_Q \frac{1}{(\tilde{Q}^2 +m^2)^2 }
 \biggr]
  =  0.02401318111946489(1) \;, \la{defP2}
\ea
where unspecified notation is explained in appendix~A.1. 
In practice $P^{ }_2$ emerges when 
considering the continuum limit, $a K \ll 1$, of the following
expression~\cite{lw}: 
\be
 \int_Q \frac{1}{\widetilde{Q}^2(\widetilde{K-Q})^2}
 \quad \stackrel{aK \ll 1}{=} \quad
 \frac{1}{(4\pi)^2} \biggl(\ln \frac{1}{a^2K^2} + 2 \biggr) + P^{ }_2
 \;.
\ee
We obtain (in the graphical notation used below, 
the circular solid line denotes the Polyakov loop to which gauge
fields can attach; blobs are the colour-electric field operators; and
wiggly lines represent gauge field propagators):
\ba
 && \hspace*{-1cm} \EleA \la{a} \\  
 \mbox{cont}: && \!\!\!
  (k_n^2 + k_i^2) \biggl\{ 1 - \frac{g^2 \Nc}{(4\pi)^2} 
  \biggl[ \frac{11}{3\epsilon} \biggr] \biggr\} \;, \nn
 \mbox{latt}: && \!\!\!
  (k_n^2 + k_i^2) \biggl\{ 1 - \frac{g^2 \Nc}{(4\pi)^2}
  \biggl[ \frac{11}{3} \ln \frac{1}{a^2\bmu^2} - \fr13 + 
  \pi^2 \biggl( 1 + \frac{20 P^{ }_1}{9} + 
  \frac{176 P^{ }_2}{3} - \frac{2}{\Nc^2} \biggr) \biggr] \biggr\} 
  \;. \nonumber
\ea

As a second set of contributions, 
we collect together effects originating
from expanding the Wilson lines between the electric fields, or in 
the denominator, to quadratic order. Here two new constants, denoted by 
$P^{ }_3$ and $p^{ }_1$ and defined in \eqs\nr{defP3} and \nr{defvP1}, 
appear:
\ba
 && \hspace*{-1cm}
 \EleB \;\quad
 \EleBB \;\quad
 \EleC \;\quad
 \EleD \;\quad
 \EleE \;\quad\;
 \EleBnorm \la{b-f} \\[3mm]
 \mbox{cont}: && \!\!\!
 \mbox{finite} + (k_n^2 + k_i^2)\, \frac{g^2 \Nc}{(4\pi)^2} 
 \biggl\{ 2 \biggl[ \frac{1}{\epsilon} + \ln \bmu^2\tau^2 \biggr] \biggr\}
 \;, 
 \nn 
 \mbox{latt}: && \!\!\!
  \mbox{finite} + 
  (k_n^2 + k_i^2) \, \frac{g^2 \Nc}{(4\pi)^2}
  \biggl\{ 2 \ln \frac{\tau^2}{a^2} + 
  \pi^2 \biggl[ 16 P^{ }_3 + 8 p^{ }_1 + 
  \frac{1}{\Nc^2}\biggl( \frac{16 P^{ }_1}{3} - 16 p^{ }_1 + \fr13
  \biggr) \biggr] \biggr\} 
  \nn 
  & & \hspace*{1.8cm} + \, k_i^2 \; \frac{g^2 \Nc}{(4\pi)^2}
  \biggl\{ \pi^2 \biggl[ \frac{8 P^{ }_1}{3} - \fr16 \biggr] \biggr\}
  \;. \nonumber
\ea
The unspecified ``finite'' terms correspond to structures which are integrable
even in the absence of an ultraviolet regulator and therefore agree
in the two schemes; they can be extracted from the continuum 
results presented in ref.~\cite{rhoE}, but are not shown here
because of their complicated appearance. 

A simple contribution, only appearing on the lattice side, concerns
a ``tadpole'' correction to the electric fields: 
\ba
 && \hspace*{-1cm}
 \EleAp \\[3mm]
 \mbox{latt}: && \!\!\!
  (k_n^2 + k_i^2) \, \frac{g^2 \Nc}{(4\pi)^2}
  \biggl\{ 
  \pi^2 \biggl[ -\frac{40 P^{ }_1}{3} + 
  \frac{16 P^{ }_1}{\Nc^2}  \biggr] \biggr\} 
  \;.  \nonumber
\ea
There are also two graphs which are finite or vanish, and therefore
yield no renormalization contribution:  
\ba
 && \hspace*{-1cm}
 \EleG \;\quad
 \EleGp \;.  % \\[3mm]
% \mbox{cont}: && \!\!\! \mbox{finite} \;, \nn 
% \mbox{latt}: && \!\!\! \mbox{finite} \;.
% \nonumber
\ea
The self-energy contribution, which is in agreement with
the classic self-energy computation in ref.~\cite{kns}, reads:
\ba
 && \hspace*{-1cm}
 \EleJ \la{j} \\[3mm]
 \mbox{cont}: && \!\!\!
  (k_n^2 + k_i^2)\, \frac{g^2 \Nc}{(4\pi)^2} 
 \biggl\{ \fr53 \biggl[ \frac{1}{\epsilon}
          + \ln \frac{\bmu^2}{K^2} \biggr] + \frac{31}{9} \biggr\} \;, 
 \nn 
 \mbox{latt}: && \!\!\!
  (k_n^2 + k_i^2) \, \frac{g^2 \Nc}{(4\pi)^2}
  \biggl\{ \fr53 \ln \frac{1}{a^2 K^2}  + \frac{28}{9} + 
  \pi^2 \biggl[1 + \frac{14 P^{ }_1}{9} +  \frac{80 P^{ }_2}{3}
   - \frac{2}{\Nc^2}\biggr] \biggr\} 
  \;. \nonumber
%%%%%%%%
\ea
The remaining graphs yield:
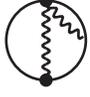
\ba
 && \hspace*{-1cm}
 \EleF \\[3mm]
 \mbox{cont}: && \!\!\!
 \mbox{finite} \;, \nn
 \mbox{latt}: && \!\!\!
 \mbox{finite} + 
  (k_n^2 + k_i^2) \, \frac{g^2 \Nc}{(4\pi)^2}
  \biggl\{ 
  \pi^2 \biggl[ 8 P^{ }_1 - 8 p^{ }_1 - \fr1{12} +  
  \frac{1}{\Nc^2}\biggl(  - {16 P^{ }_1} + 16 p^{ }_1 - \fr13
  \biggr) \biggr] \biggr\} 
  \;, \nn
%%%%%%%%
 && \hspace*{-1cm}
 \EleH \\[3mm]
 \mbox{cont}: && \!\!\!
  (k_n^2 + k_i^2)\, \frac{g^2 \Nc}{(4\pi)^2} 
 \biggl\{ -3 \biggl[ \frac{1}{\epsilon}
          + \ln \frac{\bmu^2}{K^2} +2 \biggr] \biggr\} \;, 
 \nn 
 \mbox{latt}: && \!\!\!
  (k_n^2 + k_i^2) \, \frac{g^2 \Nc}{(4\pi)^2}
  \biggl\{ -3 \biggl[ \ln \frac{1}{a^2 K^2}  + 2 \biggr] + 
  \pi^2 \biggl[ \frac{17 P^{ }_1}{3} -  48 P^{ }_2 \biggr] \biggr\} 
  \;, \nn
%%%%%%%%
 && \hspace*{-1cm}
 \EleI \la{i} \\[3mm]
 \mbox{cont}: && \!\!\!
 \mbox{finite} + 
    k_n^2 \, \frac{g^2 \Nc}{(4\pi)^2} 
 \biggl\{ \frac{3}{\epsilon}
          + \frac{13}{3}\ln \frac{\bmu^2}{K^2}
          - \frac{4}{3}\ln \frac{\bmu^2}{4 k_n^2} + \frac{14}{3} \biggr\}
 \nn 
  & & \hspace*{0.8cm} + \,  k_i^2 \, \frac{g^2 \Nc}{(4\pi)^2} 
 \biggl\{ \frac{3}{\epsilon}
          + 3 \ln \frac{\bmu^2}{K^2} + 6  \biggr\} \;, 
 \nn 
 \mbox{latt}: && \!\!\!
 \mbox{finite} + 
  k_n^2 \, \frac{g^2 \Nc}{(4\pi)^2}
  \biggl\{ \fr{13}3 \ln \frac{1}{a^2 K^2}  -\fr43 \ln\frac{1}{4 a^2 k_n^2}
  + \frac{14}{3} + 
  \pi^2 \bigl[ - 3 P^{ }_1 +  48 P^{ }_2  \bigr] \biggr\} 
 \nn 
  & & \hspace*{0.8cm} + \, k_i^2 \, \frac{g^2 \Nc}{(4\pi)^2}
  \biggl\{ 3 \ln \frac{1}{a^2 K^2} + 6 + 
  \pi^2 \biggl[ - \frac{17 P^{ }_1}{3}  +  48 P^{ }_2  \biggr] \biggr\} 
  \;. \nonumber
\ea
The last contribution turns out to be quite tedious to extract, but 
given its simple final appearance we refrain from 
dwelling on more details here. 

%%%%%%%%%%%%%%%%%%%%%%%%%%% SECTION %%%%%%%%%%%%%%%%%%%%%%%%%%%%%%%%%%
%
\section{Determination of the renormalization factor}
\la{se:ZE}

Summing together \eqs\nr{a}--\nr{i}
all appearances of the ``three-dimensional'' lattice 
constant $p^{ }_1$, defined in \eq\nr{defvP1}, cancel. We are left with 
\ba
  \mbox{cont}: && \!\!\!
   k_n^2\, \biggl\{ 1 + \frac{g^2 \Nc}{(4\pi)^2} 
  \biggl[ \mbox{finite} 
   - \fr43 \ln\frac{\bmu^2}{4k_n^2} 
   + 3 \ln\frac{\bmu^2}{K^2} 
   + 2 \ln (\bmu^2\tau^2) + \fr{19}9  \biggr] \biggr\} \nn
   && \hspace*{-5mm} + \, 
   k_i^2\, \biggl\{ 1 + \frac{g^2 \Nc}{(4\pi)^2} 
  \biggl[  \mbox{finite} + 
     \fr53 \ln\frac{\bmu^2}{K^2} + 2 \ln (\bmu^2\tau^2) + \frac{31}{9}  
  \biggr] \biggr\} \;, \\
 \mbox{latt}: && \!\!\!
   k_n^2\, \biggl\{ 1 + \frac{g^2 \Nc}{(4\pi)^2} 
  \biggl[  \mbox{finite} 
   - \fr43 \ln\frac{\bmu^2}{4k_n^2} 
   + 3 \ln\frac{\bmu^2}{K^2} 
   + 2 \ln (\bmu^2\tau^2) + \fr{19}{9} 
  \nn && \hspace*{2cm} + \, \pi^2 \biggl( 
     -\fr1{12} - \frac{10 P^{ }_1}{3}
  + \frac{16 P^{ }_1}{3 \Nc^2} -32 P^{ }_2 + 16 P^{ }_3
  \biggr)  
  \biggr] \biggr\} \nn
   && \hspace*{-5mm} + \, 
   k_i^2\, \biggl\{ 1 + \frac{g^2 \Nc}{(4\pi)^2} 
  \biggl[  \mbox{finite} + 
     \fr53 \ln\frac{\bmu^2}{K^2} + 2 \ln (\bmu^2\tau^2) + \frac{31}{9}
  \nn && \hspace*{2cm} + \, \pi^2 \biggl( 
     -\fr1{4} - \frac{10 P^{ }_1}{3}
 + \frac{16 P^{ }_1}{3 \Nc^2} -32 P^{ }_2 + 16 P^{ }_3
  \biggr)  
 \biggr] \biggr\} 
  \;. \la{latt_sum}
\ea
Subtracting the two and noting that in the remaining terms 
the sum-integration measure in \eq\nr{measure} implies that 
we can substitute $k_n^2\to \frac{D-1}{D-2}(k_n^2+k_i^2)$, 
$k_i^2 \to \frac{-1}{D-2}(k_n^2 + k_i^2)$ 
with the price of inessential contact terms, 
we obtain (here also $D\to 4$)
\ba
 \mbox{latt} - \mbox{cont} = (k_n^2 + k_i^2)
 \biggl\{ 0 + \frac{g^2 \Nc}{(4\pi)^2} 
 \biggl[ 
  \pi^2 \biggl(  - \frac{10 P^{ }_1}{3} + \frac{16 P^{ }_1}{3 \Nc^2}
  - 32 P^{ }_2 + 16 P^{ }_3 \biggr) 
 \biggr]
 \biggr\}
 \;. \la{final_diff}
\ea
This difference needs to be cancelled by a renormalization
factor $\mathcal{Z}_\rmii{E}$ in order for the lattice and continuum
results to agree: 
\be
 G^{ }_{\rmii{E},\rmi{cont}}(\tau)  \equiv 
 \mathcal{Z}^{ }_\rmii{E} \, 
 G^{ }_{\rmii{E},\rmi{latt}}(\tau) 
 \;. \la{ZE_def}
\ee
Eq.~\nr{final_diff} leads to the next-to-leading order perturbative
expression for $\mathcal{Z}^{ }_\rmii{E}$:
\ba
 \mathcal{Z}^{ }_\rmii{E} 
 & = &  
 1 +  \frac{g^2 \Nc}{16} 
  \biggl( \frac{10 P^{ }_1}{3} - \frac{16 P^{ }_1}{3 \Nc^2}
  + 32 P^{ }_2 - 16 P^{ }_3 \biggr) 
 + \rmO(g^4)
 \nn 
 & = & 1 + \frac{2 g^2 \CF P^{ }_1}{3}
%  + \frac{g^2 \Nc}{8} 
%  \biggl( - P^{ }_1 
%  + 16 P^{ }_2 - 8 P^{ }_3 \biggr) 
 + \rmO(g^4) 
 \;. \la{final}
\ea
In the last step we made use of the non-trivial relation 
determined in \eq\nr{magic}.
This simple expression, obtained through a substantial effort, 
constitutes our final result. 

In ref.~\cite{kappaE}, a provisional result 
for $\mathcal{Z}^{ }_\rmii{E}$ was given, 
based on a lattice HQET computation
of the type in ref.~\cite{ee1}, concerning
the renormalization of the spatial components of the Noether
current. Our result in \eq\nr{final} agrees with the part
proportional to $\CF P^{ }_1$ in the estimate of ref.~\cite{kappaE}, 
however that result has a large additional term 
proportional to $-\CF p^{ }_1$; 
this structure is the same that appears 
in connection with the mass renormalization
of a static quark. We are not surprised by the difference, given 
that $G^{ }_\rmii{E}$ does not directly correlate the spatial
components of the Noether current but their time derivatives. 
Taking time derivatives involves a specific discretization, 
which interferes with the movement of the static quark of mass 
correction $\sim \CF p^{ }_1/a$ by a distance $\sim a$ in the time direction. 
It is non-trivial to account for such effects, 
and conceivable that something got lost
in the HQET estimate. 

Inserting a numerical value for the coefficient 
$P^{ }_1$ (cf.\ \eq\nr{defP1}) and $\Nc = 3$ for the 
number of colours, and using \eq\nr{L_MS} to re-express $g^2$ through $g_0^2$, 
we obtain
\be
 \mathcal{Z}^{ }_\rmii{E} 
 \; = \; 
 1 + g_0^2 \times  
  0.13771856909427574(1)
 + \rmO(g_0^4) 
 \;. \la{final_num}
\ee
Recalling $g_0^2 = 6/\beta^{ }_0$, where $\beta^{ }_0$ 
is the bare lattice coupling, and noting that values in the range
$\beta^{ }_0 \sim 7...8$ have been used in simulations
attempting to make a continuum extrapolation~\cite{kappa}, 
a rather modest $\sim 12\%$ renormalization effect is found. This
can be compared, for instance, with the renormalization factor
associated with a local version of the vector current: in that
case the $\rmO(g_0^2)$ correction~\cite{Zp1,Zp2,Zp3} is slightly
larger than ours,\footnote{%
 Note however that $Z^{ }_\rmii{V}$ is associated with a single operator, 
 so that in the 2-point function $Z^{2}_\rmii{V}$ appears; the 
 correction in this quantity is then twice as large as in our
 $\mathcal{Z}^{ }_\rmii{E}$, which concerns the full correlator.  
 } 
but nevertheless it dominates the full result
including all $\rmO(g_0^4)$ effects~\cite{ZV1,ZV2}.
Therefore we may expect a $\sim 5$\%
theoretical uncertainty from effects of $\rmO(g_0^4)$ in the present case.  

Finally we note that in ref.~\cite{kappa} a provisional version 
of \eq\nr{final_num} was used in connection with a lattice analysis. 
Unfortunately that version contained an algebraic error, whereby
the numerical value was estimated as 
$
 \mathcal{Z}^{ }_\rmii{E} 
  \simeq 
 1 + g_0^2 \times 0.079 
$.
For $\beta_0 \sim 7 ... 8$
the difference with respect to \eq\nr{final_num} is of a similar magnitude
as our estimate for $\rmO(g_0^4)$ corrections. However, 
given that systematic errors were estimated 
generously in ref.~\cite{kappa} and that they are predominantly associated
with analytic continuation from the imaginary-time correlator to 
a spectral function, the final results should 
not be significantly affected within their 50\% uncertainties. 
(Note in particular that one of the ans\"atze considered in 
ref.~\cite{kappa}, denoted by $3a$, left the overall normalization
of the ultraviolet part of the spectral function open, which 
roughly corresponds to leaving $\mathcal{Z}^{ }_\rmii{E}$ open.)

%%%%%%%%%%%%%%%%%%%%%%%%%%% SECTION %%%%%%%%%%%%%%%%%%%%%%%%%%%%%%%%%%
%
\section{Conclusions and outlook}
\la{se:concl}

The purpose of this technical contribution has been to report on 
the result of a 1-loop computation in lattice perturbation theory
for the renormalization factor defined in \eq\nr{ZE_def}. The result
for the case that the action is discretized according to 
Wilson's classic prescription and the observable is discretized 
as specified in \eq\nr{GE_lattice} %; and we set $\Nc = 3$ and  $\Nf = 0$, 
is given in \eq\nr{final_num}. The numerical magnitude of this 
correction is rather modest, and suggests that future updates of  
simulations such as those in ref.~\cite{kappa} can  
attempt approaching the continuum limit, once the results are 
multiplied by the perturbative renormalization 
factor that we have determined.    

For a definitive lattice result, renormalization
should be promoted to a non-perturbative level. 
One possibility for this is to make use of Yang-Mills
gradient flow~\cite{wilson}: 
following ref.~\cite{gradient}, we expect that if the electric fields 
and Wilson lines between them are computed from gauge fields 
at a positive flow time $t > 0$, then the continuum limit can be 
taken at each $t$ and leads to a finite regularization-independent
result. Subsequently the limit $t\to 0^+$ needs to be taken, 
but this should not lead to any divergences 
since $G^{ }_{\rmii{E},\rmi{cont}}$ is finite~\cite{eucl,rhoE}. 
A program of a similar type has been suggested for the energy-momentum 
tensor some time ago~\cite{hs1,ap}, 
and numerical tests have been carried out in the meanwhile~\cite{hs2}. 
Recently gradient flow has also been 
used for defining renormalized Polyakov 
loop expectation values~\cite{ps,sd}. 

Ultimately, for physical QCD, dynamical quarks
need to be included in the analysis. The discretization 
choices made in this context are of a great variety, 
so computations within lattice perturbation theory 
become cumbersome and non-universal. Therefore we have not embarked on 
the inclusion
of dynamical quarks; this is probably sensible anyways only in combination 
with non-perturbative renormalization. 

%%%%%%%%%%%%%%%%%%%%%%%%%%% SECTION %%%%%%%%%%%%%%%%%%%%%%%%%%%%%%%%%%
%
\section*{Acknowledgements}

We thank J.~Langelage for collaboration at initial stages of this 
project. 
The work was supported by the Swiss National Science Foundation
(SNF) under grant 200020-155935.

%%%%%%%%%%%%%%%%%%%%%%%%%%% APPENDIX %%%%%%%%%%%%%%%%%%%%%%%%%%%%%%%%
%
\appendix
\renewcommand{\thesection}{Appendix~\Alph{section}}
\renewcommand{\thesubsection}{\Alph{section}.\arabic{subsection}}
\renewcommand{\theequation}{\Alph{section}.\arabic{equation}}

%%%%%%%%%%%%%%%%%%%%%%%%%%% SECTION %%%%%%%%%%%%%%%%%%%%%%%%%%%%%%%%%%
%
\section{Technical ingredients}

%%%%%%%%%%%%%%%%%%%%%%%%%%% SUBSECTION %%%%%%%%%%%%%%%%%%%%%%%%%%%%%%%
%
\subsection{Basic notation}

Lattice momenta are denoted by 
\be
 \tilde{q}_\mu \equiv \frac{2}{a} \sin \frac{a q_\mu}{2}
 \;, \quad
 \undertilde{q_\mu} \,\equiv\, \cos \frac{a q_\mu}{2}
 \;,
\ee
where $Q = (q_n,\vec{q})$ and 
$a$ is the lattice spacing. 
The sum-integration measure stands for 
\be
 \Tint{Q} \equiv T \sum_{q_n} \int_{-\frac{\pi}{a}}^{\frac{\pi}{a}}\!
 \frac{{\rm d}^3 \vc{q}}{(2\pi)^3}
 \;, \la{measure2}
\ee
where the Matsubara frequencies read 
\be
 q_n = 2 \pi n T = \frac{2\pi n}{a N_\tau}
 \;, \quad 
 n = 1, \ldots, N_\tau
 \;, 
\ee
with $N_\tau$ denoting the temporal extent in lattice units. 
In the zero-temperature limit we denote $q_n \to q_0$, 
and the measure becomes
\be
 \int_Q 
 \equiv  
 \int_{-\frac{\pi}{a}}^{\frac{\pi}{a}}\!
 \frac{{\rm d} q_0}{2\pi} 
 \int_{-\frac{\pi}{a}}^{\frac{\pi}{a}}\!
 \frac{{\rm d}^3 \vc{q}}{(2\pi)^3}
 \;. 
\ee
In the renormalization
factors we can always 
use zero-temperature measures, given that thermal effects
are finite and independent of the regularization scheme. 

%%%%%%%%%%%%%%%%%%%%%%%%%%% SUBSECTION %%%%%%%%%%%%%%%%%%%%%%%%%%%%%%%
%
\subsection{Lattice constants}

Apart from the ``four-dimensional'' constants 
$P^{ }_1$ and $P^{ }_2$ defined in 
\eqs\nr{defP1} and \nr{defP2} and 
from an analogous ``three-dimensional'' constant $p^{ }_1$
defined in \eq\nr{defvP1}, a ``mixed''
constant also makes an appearance in the analysis:
\be
 P^{ }_3 \equiv \lim_{m\to 0}
 \biggl[
   \frac{2 \ln ( {a^2 m^2} ) }{(4\pi)^2}  + 
   \int_Q \frac{1}{(\tilde{q}^2 + m^2)(\tilde{Q}^2 + m^2)} 
 \biggr]
 \;, \la{defP3} 
\ee
where
$
 \tilde{q}^2 \equiv \sum_{i=1}^3 \tilde{q}_i^2
$
and
$
 \tilde{Q}^2 \equiv \tilde{q}_0^2 + \tilde{q}^2
$. 
This plays a role in the sum-integral
\be
 I(\tau) \;\equiv\; 
 \Tint{Q} \frac{e^{i q_n\tau} - 1 }{\tilde{q}^2\tilde{Q}^2}
 \;, 
\ee
which in dimensional regularization reads
\be
 \left. I(\tau) \right|_\rmi{cont} = 
 - \frac{2}{(4\pi)^2}
 \biggl( \frac{1}{\epsilon} + \ln \frac{\bmu^2\tau^2}{4} + 2 + 2\gammaE
 \biggr)
 + 
 \int_0^\infty \! {\rm d}q \, \frac{ \nB{}(q) 
 \, \bigl[ e^{q \tau} + e^{- q \tau} - 2 \bigr] }{ (2\pi)^2 q}
 \,+\, \rmO(\epsilon) \;, 
\ee
where $\nB{}(q) \equiv 1/[\exp(\beta q) - 1]$ is the Bose distribution. 
On the lattice it evaluates to
\be
 \left. I(\tau) \right|_\rmi{latt} = 
 - \frac{2}{(4\pi)^2}
 \biggl( \ln \frac{\tau^2}{4a^2 } + 2 + 2\gammaE
 \biggr) - P^{ }_3 
 \,+
 \int_0^\infty \!\! {\rm d}q \, \frac{ \nB{}(q) 
 \, \bigl[ e^{q \tau} + e^{- q \tau} - 2 \bigr] }{ (2\pi)^2 q}
 \,+\, \rmO\Bigl(\frac{a^2}{\tau^2}\Bigr)
 \;. 
\ee
This sum-integral appears (only) in the 1st and 2nd diagrams of 
\eq\nr{b-f}. 

Remarkably, $P^{ }_3$ as defined in \eq\nr{defP3} is related to 
the constants
$P^{ }_1$ and $P^{ }_2$ as defined in \eqs\nr{defP1} and \nr{defP2}.
One way to see this is to represent the propagators as
$
 1/\Delta = \int_0^\infty \! {\rm d}t \, e^{-t \Delta}
$
and then to carry out the angular integrals as 
$
 \int_{-\pi}^{\pi} \! \frac{{\rm d}\theta}{2\pi} \, e^{t\cos\theta} = 
 I^{ }_0(t) 
$,
where $I_0$ is a modified Bessel function. For $P^{ }_2$ and $P^{ }_3$
two variables $t_1,t_2$ may be introduced for the two propagators, 
but we can subsequently 
substitute variables as $x \equiv t_1 + t_2$ and $y \equiv t_1$. The 
expression for $P^{ }_3$ contains the integral
\be
 \int_0^x \! {\rm d}y \, e^{-y} I^{ }_0(y) 
 = x e^{-x} \, \bigl[I^{ }_0 (x) + I^{ }_1(x) \bigr]
 \;. 
\ee
Inserting $I^{ }_1(x) = I'_0(x)$ yields 
\ba
 8( 2 P^{ }_2 - P^{ }_3) - P_1
 & = & 
 \fr12 \int_0^\infty \! {\rm d}x \, e^{-4 x} \, 
 I_0^3(x) \, 
 \Bigl[ 
  4 x I^{ }_0(x) - 4 x I'_0(x) - I^{ }_0(x) 
 \Bigr]
 \nn 
 & = & 
 - \fr12 \int_0^\infty \! {\rm d}x \,
 \frac{{\rm d}}{{\rm d}x} 
 \Bigl[ x e^{-4 x} I_0^4(x) \Bigr]
 \; = \; 0
 \;. \la{magic}
\ea

Apart from the coefficients $P^{ }_1,P^{ }_2$ and $P^{ }_3$ defined in 
\eqs\nr{defP1}, \nr{defP2} and \nr{defP3}, the three-dimensional
lattice integral~\cite{math1,math2}
\be
  p^{ }_1  \equiv  \int_{\vec{q}} \frac{a}{\tilde{q}^2}
  =  \Gamma^2[\fr1{24}]\Gamma^2[\fr{11}{24}]
 \frac{\sqrt{3}-1}{ 192 \pi^3}  \;, \la{defvP1} 
 \quad
 \int_{\vec{q}} = \int_{-\frac{\pi}{a}}^{\frac{\pi}{a}}\!
 \frac{{\rm d}^3 \vc{q}}{(2\pi)^3}
 \;, 
\ee
also appears in the results for the individual graphs.
It cancels in the sum, cf.\ \eq\nr{latt_sum}.

%%%%%%%%%%%%%%%%%%%%%%%%%%% SUBSECTION %%%%%%%%%%%%%%%%%%%%%%%%%%%%%%%
%
\subsection{Frequently needed relations}

There are many identities between various lattice integrals that are
helpful for simplifying the computation; a collection sufficient for 
our purposes can be found in appendix~B of ref.~\cite{lw}. Let us here 
just recall two examples that appear particularly often: 
\be
 \int_Q \frac{a^4 \tilde{q}_\mu^2 \tilde{q}_\nu^2 }{\tilde{Q}^4}
  = \frac{P^{ }_1}{3} + 
 \delta^{ }_{\mu\nu} \biggl(\fr14 - \frac{4 P^{ }_1}{3} \biggr) 
 \;, \quad
 \int_\vc{q} \frac{a^3 \tilde{q}_i^2 \tilde{q}_j^2 }{\tilde{q}^4} 
  = \frac{p^{ }_1}{3} + \delta^{ }_{ij}
 \biggl(\fr13 - p^{ }_1 \biggr)
 \;. 
\ee

%%%%%%%%%%%%%%%%%%%%%%%%%%% SUBSECTION %%%%%%%%%%%%%%%%%%%%%%%%%%%%%%%
%
\subsection{Thermal sums}

A basic thermal sum appearing is 
\be
%  T \sum_{p_n'} \frac{e^{i p_n\tau} - 1}{\tilde{p}_n^2 + \omega^2}
% = 
  T \sum_{q_n} \frac{e^{i q_n\tau} }{\tilde{q}_n^2 + \omega^2}
 = 
  \frac{a \bigl[ e^{k x} + e^{(N_\tau - k)x} \bigr]}
  {2 \bigl[ e^{N_\tau x } - 1\bigr] \sinh x  }
 \;, \quad
 x = 2 \mathop{\mbox{asinh}} \frac{a\omega }{2}
 \;, \quad
 \frac{k}{N_\tau}   % =  \frac{ak}{aN_\tau}
  = \frac{\tau }{\beta}
 \;. \la{sum}
\ee
Methods for carrying out sums of this type were discussed
in ref.~\cite{sum}.
{}From this sum others can be obtained through the limits $\omega\to 0$
and/or $\tau\to 0$ and through the omission of the Matsubara zero mode, 
in which case we denote the variable by $q_n'$. In particular, 
\ba
 T \sum_{q_n'} \frac{e^{i q_n \tau} - 1 }{\tilde{q}_n^2} & = &  
 \frac{T \tau(\tau - \beta)}{2}  \;, \\[1mm] 
 T \sum_{q_n'} \frac{ 1 }{\tilde{q}_n^2} & = &  
 \frac{T (\beta^2 - a^2)}{12}  \;, \\[1mm] 
 \numerator \hspace*{7mm}
 & = & 
 \frac{T (2 \tau - \beta)}{2i} 
 \;.
\ea

%%%%%%%%%%%%%%%%%%%%%%%%%%%%%%%%%%%%%%%%%%%%%%%%%%%%%%%%%%%%%%%%%%%%%%%%%%%

\end{document}